Corresponding Author: Dr. Nicolas A. Pereyra, Ph.D.

Corresponding Author's Institution: Prism Computational Sciences, Inc.

First Author: Nicolas A Pereyra, PhD

Order of Authors: Nicolas A Pereyra, PhD; Joseph J MacFarlane, PhD; Pamela R Woodruff, PhD; Igor E Golovkin, PhD; Ping Wang, PhD

Abstract:

# Short-N characteristic radiative transfer method in 2D cylindrical RZ coordinates

Nicolas A. Pereyra*, Joseph J. MacFarlane, Pamela R. Woodruff, Igor E. Golovkin, and Ping Wang

Prism Computational Sciences, 455 Science Drive, Suite 140, Madison, WI 53711 USA

**Abstract**

We present a new radiative transfer method "Short-N Characteristics" that is a hybrid between the standard long characteristic methods and the standard short characteristic methods. We have implemented the numerical method within the *SPECT3D* imaging and spectral analysis application. Numerical experiments have shown that the short-N method is capable of reproducing the accuracy of the long characteristic methods, while delivering the CPU time efficiency of the short characteristic methods. We apply the short-N method to 2.5D cylindrical coordinates, i.e., we assume a 3D problem under azimuthal symmetry leading to a 2D mathematical problem with $\rho$-z as the independent spatial coordinates (2DRZ). We are currently working to extend this new method to other geometries.

*Keywords*: 2.5D Plasma Simulation; Collisional Radiative Kinetics; Cylindrical Coordinates; Numerical Method; Plasma Spectroscopy

* Corresponding author: Fax 1 608 268 9180
  E-mail address: pereyra@prism-cs.com (Nicolas A. Pereyra)

## 1. Introduction

Starting from plasma hydrocode data, *SPECT3D* performs detailed spectral and population calculations in one, two and three dimensions for various geometries, computing line of sight images, spectral and radiative properties. That is, it post processes the hydrodynamic data by self-consistently calculating the atomic population levels and the radiative properties throughout the plasma.

The radiative properties within a given volume element of the plasma, as well as the radiative properties measured by detectors outside the plasma, may depend on the interaction of the radiation generated throughout the plasma with all the volume elements of the plasma. Therefore radiative transfer calculations are essential in the determination of atomic level populations and radiative properties.

*SPECT3D* has been applied to a variety of physical systems:

- The study of dynamic hohlraum radiation and implosion physics (e.g., Epstein et al. [1]; Bailey et al. [2]; Bailey et al. [3]; Golovkin et al. [4]; MacFarlane et al. [5]; Sinars [6]);
- The study of x-ray photoionization of plasmas (e.g., Cohen et al. [7]; Shupe et al. [8]);
- The study of hohlraum radiation driven jets (e.g., Lawrence et al. [9]; Lawrence et al. [10]).
- The study of laser produced and discharge produced EUVL radiation sources (e.g., Rettig et al. [11]; MacFarlane et al. [12]; MacFarlane et al. [13])
- The study of short-pulse-laser experiments aimed towards fast ignition applications (e.g., MacFarlane et al. [14])
- The astrophysical study of x-ray emission and absorption in hot star winds (e.g., MacFarlane et al. [15]; Abing et al. [16]; Johnson et al. [17])

*SPECT3D* achieves its results through a numerical iterative process. It initially starts with LTE atomic level populations and does a detailed radiative transfer calculation. From the radiative transfer results, the mean intensity is computed at each spatial grid point of the plasma and the corresponding atomic radiative rates. From these computed mean intensities and atomic radiative rates, the corresponding atomic population levels are calculated at each spatial grid point. The new population levels are used to recompute the detailed radiative transfer and new mean intensities are computed. The process continues until the population levels converge or until a sufficiently large number of iterations has been done. Thus, the task of coupling the detailed atomic data consistently with the radiative transfer calculations and with the hydrodynamic data to obtain the radiative properties of the plasma, becomes a numerically intensive one, particularly when the calculations are done in two or three dimensions.

To solve the radiative transfer equations in 2D cylindrical RZ (2DRZ) geometry, *SPECT3D* previously implemented the long characteristic method. That is, a 3D angle-direction-grid is defined with an optimized numerical quadrature for the calculation of mean intensity. For each volume element of the plasma and for each angle, a ray is considered that passes through the given volume element at the given angle and that extends throughout the plasma. The radiative transfer equation is then self-consistently solved throughout all the volume elements intersected by the ray, and the intensity for the given volume element and the given angle is obtained. The process is repeated, and the intensities of all volume elements at all angles are obtained. The radiative transfer equation is solved in this manner for each ray generated by each angle and by each volume element, significantly increasing the required CPU time for the post processing of a hydrodynamic simulation.

The aim of the efforts outlined here is to significantly reduce the CPU time required by the 2DRZ calculations, without significantly reducing the accuracy of the results, by applying the short-N radiative transfer method that we describe below. For comparisons we use the CPU times required for post processing a hydrodynamic simulation of a 2D-cylindrical plastic shell filled with Ar-doped deuterium (0.05% Ar). With the long characteristic methods for spatial grids of the order of 100x100 volume elements and a detailed atomic model, the calculation currently takes over two weeks to complete on a single CPU system.

It should be noted that we have previously increased the clock-time efficiency through parallelization (Golovkin [18]; Golovkin et al. [19]), and that the clock-time required by *SPECT3D* is significantly reduced if it runs on a cluster (the clock-time efficiency increasing with the size of the cluster).

The efforts described here will significantly reduce the clock-time required for post processing plasma hydrodynamic simulations both for clusters and for single CPU systems.

The results that are shown here, are computed from the implementations of the radiative transfer methods within the *SPECT3D* application.

**2. Short-N Characteristics Method and 2DRZ Geometry**

In the 2DRZ geometry, all physical parameters are expressed in terms of cylindrical coordinates (ρ-φ-z) and are assumed to be independent of φ. That is, azimuthal symmetry is assumed. Spatial variations of physical parameters are thus reduced to the dependence on ρ-z coordinates.

Physically, however, the plasma is in three-dimensional space, and thus the radiative transfer calculations must be done in a three-dimensional spatial grid. The corresponding three-dimensional spatial grid is a series of toriods obtained by rotating the initial 2D ρ-z spatial grid about the axis ρ = 0. A significant consequence of generating a three dimensional grid from 2DRZ geometry (for radiative transfer calculations), is that a ray that passes through the plasma, intersecting several different volume elements, will have different angles relative to each different volume element.

In applying the long characteristic method for radiative transfer, all rays passing through each volume element at each angle are considered. As discussed above, this leads to a numerically intensive calculation. This in turn leads us to consider a short characteristic method in order to reduce the CPU-time requirements for the radiative transfer calculations. Short characteristic methods have been discussed in detail in the literature (see Noort et al. [20]. and references therein).

The basic idea of short characteristic methods is to reduce the radiative transfer calculations to a series of "short" spatial segments (rather than through "long" segments that extend throughout the plasma as in the long characteristic methods). Each short segment is contained within a single volume element and its beginning and ending points are at the walls of the volume element. The direction of the short segment is set by the numerical angular grid of the radiative transfer calculation. The intensity at the beginning of the segment is set by the corresponding intensities of the adjacent volume element (or by the boundary conditions of the plasma if the volume element under consideration is at the spatial boundary of the plasma). For each volume element and for each angle, the radiative transfer computations needed are done along the single volume element with the boundary conditions given by the previous volume element. This significantly reduces the amount of required radiative transfer computations, and thus the required CPU time.

During the development of this work we found that the accuracy of the short characteristic methods could be increased, while still significantly reducing the CPU time, by extending the radiative transfer calculations over "N" volume cells, where N is a small integer (rather than fixing N=1 as in the standard short characteristic method). We are currently using a default value of 3 for "N", although this parameter may be set to an arbitrary nonzero integer. We denote this approach as the "Short-N Characteristic Method". In Figure 1 we show a simplified schematic representation of the method. Below we show results obtained for different N.

In addition to the gain in CPU time efficiency, the short-N characteristic method also has the advantage of less memory requirements compared with the long characteristic method. For the calculations with the highest spatial resolution with the most detailed atomic model considered here (spatial grid of 150x80; 404 atomic energy levels for Ar), the short-N method with N=3 required ~580MB of RAM, while the same model with the long characteristic method required ~890MB of RAM.

## 3. Geometric Calculations

Since *SPECT3D* post processes hydrodynamic data with arbitrary spatial grids, geometric calculations are implemented after the hydro data is read and before the actual radiative transfer computations. The geometric calculations are independent of the values of intensities and in particular of the photon energies considered. The results from these computations are stored in memory and accessed as needed during the radiative transfer calculations. Hence the geometric computations do not significantly increase the CPU-time required for post processing hydrodynamic data.

In these computations, geometric parameters for each volume element and for each angle (such as intersection positions of short segments with volume element walls, relative angles, etc.) are calculated and stored. The required interpolation parameters for intensities, as well as the corresponding dependencies with respect to the short-N characteristic method are also determined and stored. The order of the different short-N segments, on which the radiative transfer computations are done, is also determined at this point to ensure that corresponding boundary conditions for each short segment will have been computed before the radiative transfer computations are done on the given segment.

## 4. Radiative Transfer Computations

Radiative transfer computations are performed on the short-N segments in the order previously determined by the geometric calculations. First, radiative transfer computations are done on short-N segments that have a beginning point at the boundary of the plasma, followed by computations in adjacent volume elements in inward directions. The process continues, eventually computing on short-N segments in the outward directions until every short-N segment has been computed upon.

Mean intensities are then calculated for each volume element, and the general numerical iterative process described above proceeds until atomic level populations converge.

## 5. CPU-time efficiency: Short-N Characteristics vs. Long Characteristics

The gain in CPU-time-efficiency of the short-N radiative transfer method vs. the long radiative method will depend on the size of the spatial grid. For a spatial grid of n×n the respective CPU-times approximately vary with 'n' as

$$\text{Long-Characteristic (Radiative-Transfer-Time)} \sim n^3$$

$$\text{Short-N-Characteristic (Radiative-Transfer-Time)} \sim n^2$$

That is, short-N characteristic methods will be of the order of 'n' times more efficient that the long-characteristic method. The larger the spatial grid, the larger the relative efficiency of the short-N method will be.

Now, the total CPU-time for post processing a hydro data file is dominated by two sets of computations: first, the radiative transfer itself, which is the focus of the efforts described here, and second, the calculation of the opacities used in the radiative transfer. In turn, the opacity calculation time is proportional to the number of volume elements in the spatial grid, that is

$$\text{Opacity Calculation Time} \sim n^2$$

Thus, the relative gain in CPU-time efficiency with the short-N method will increase for higher spatial resolutions and will also increase for less-detailed atomic models that require less time for the opacity calculations.

As we show below, post processing a high spatial resolution hydro data file with detailed atomic models that took several weeks on a single CPU system with the long-characteristic method, can now be done within 36 hours on the same computer system with the short-N method.

## 6. Comparison with Analytic Results

As with the earlier long characteristic method implemented in *SPECT3D*, the short-N characteristic method has been tested against analytic results. In particular, we consider a plasma, with a uniform source function $S_v$, occupying a cylindrical volume of height H and radius R, represented with a spatial grid of 10×20 ($\rho \times z$).

In Figure 2 we show the plots of the mean intensity vs. radius for optical depths, $\tau_R$, along the radial direction of (above) $\tau_R$=1.825 and (below) $\tau_R$=0.165, for the value of the short-N-parameter of N=3, superimposed on the analytic solutions. We see from Figure 2 that the differences between the numerical short-N mean intensities and the analytic mean intensities are larger towards the center of the cylinder (r=0, z=0). This is because the intensities in these regions, in all directions, are obtained through the radiative transfer calculations over many short-N segments (rather than through a single ray as in the long characteristic methods).

As a consistency check, in Figure 3 we show plots of the mean intensity vs. radius for optical depths along the radial direction of (above) $\tau_R$=1.825 and (below) $\tau_R$=0.165, for the value of the short-N-parameter of N=20. As expected, the agreement with the analytic solution is excellent.

## 7. Results and Discussion – Ar Doped Problem

The results shown in this section correspond to the same physical problem: A spherical plastic shell with radius 0.028 cm and thickness of 0.002 cm filled with Ar-doped deuterium (0.05% Ar). The temperature of the Ar-doped deuterium varies from 1500 eV at the plasma center to 200 eV at the plastic shell. The density of the Ar-doped deuterium varies from 0.3 g cm$^{-3}$ at the plasma center to 0.5 g cm$^{-3}$ at the plastic shell. These physical parameters correspond to an idealized model for a typical ICF (Inertial Confinement Fusion) experiment at the time of implosion stagnation.

Azimuthal symmetry is assumed and all cases are computed under 2DRZ geometry. What varies in the different cases of this section is the spatial resolution implemented and the atomic model used for Ar in the calculations.

In Figure 4 we show plots of the spectra produced by the long characteristic method and the short-N method for different values of the N-parameter for a low spatial resolution of 10x10 and a simple atomic model for Ar of 11 energy levels. In Figure 5 we show plots of the spectra produced by the long characteristic method and the short-N method for N=1 and N=2, for a low spatial resolution of 10x10 and a detailed atomic model for Ar of 404 energy levels. Note that the agreement with the long characteristic method tends to increase with larger N, illustrating that the accuracy of the short-N method increases with the N-parameter.

In Table 1, we show the comparison of CPU times between the short-N characteristic method and the long characteristic method applied to the Ar Doped problem for a mid-spatial resolution of 50x50 and a simple atomic model for Ar of 11 energy levels. In Table 2, we show the equivalent results for a detailed atomic model for Ar of 404 energy levels. By comparing the fourth and sixth columns of Tables 1 and 2, we note that the relative gain in efficiency for the complete computation is less than the relative gain in efficiency of the radiative transfer calculation. This is because a significant part of the computations are for opacity calculations that are performed previous to the radiative transfer calculations. The effects on the CPU-time due to the opacity calculations can be clearly seen by comparing the fourth and sixth columns of Tables 1 and 2. One can see that, although the relative gain in efficiency in the radiative transfer calculations is similar for both tables, the relative gain for the total computation is considerably higher for the simpler atomic models (Table 1) since it requires significantly fewer opacity calculations.

In Figure 6 we show plots of the spectra produced by the long characteristic method and the short-N method for different values of the N-parameter for a high spatial resolution of 150x80 and a simple atomic model for Ar of 11 (eleven) energy levels, and find conclusions similar to those of the low spatial resolution model of Figures 4 and 5.

In Figure 7 we plot the ratio of the total CPU times for the calculation using the long characteristic method and the short-N characteristic method for N=1, 2, and 3 vs. the square root of the total number of volume elements in the spatial grid. For this figure the simple atomic model of 11 levels for Ar is used. In Figure 7, we see that significant increases in the time efficiency of the radiative transfer calculations are gained with the short-N characteristic method with respect to the long characteristic method. We also see that the relative gain in efficiency increases with higher spatial resolution, approximately in proportion to the number of volume elements along a single dimension.

The most "realistic" computation considered here, the one with the high spatial resolution of 150x80 and a detailed atomic model of 404 energy levels for Ar, shows a gain in CPU-time efficiency for the total computation of over a factor of ten. This run that previously took several weeks with the long characteristic method, is now running within 36 hours with the short-N method without loss of accuracy.

## 8. Summary and Conclusions

- We have developed a new radiative transfer method named "Short-N Characteristic", and have implemented it in the *SPECT3D* plasma simulation application for 2DRZ geometries. The method is a hybrid between the standard long characteristic methods and the standard short characteristic methods.
- The short-N radiative transfer method has the advantage of simultaneously keeping the accuracy of the long characteristic methods and the CPU-time efficiency of the standard short characteristic methods.
- The implementation of the method has been successfully tested against analytical solutions in 2DRZ geometries.
- In one of the benchmarking tests, for a high spatial resolution of 150x80 with a detailed atomic model of 404 energy levels, a run that previously took several weeks with the long characteristic method, is now running within 36 hours with the short-N method without loss of accuracy.
- We are currently working on implementing the short-N characteristic method for other geometries.

## 9. Acknowledgements

We would like to acknowledge the financial support of Sandia National Laboratories and the financial support of the U.S. Department of Energy for the development of this work.

**Figure Captions**

Fig 1. Simplified schematic illustration of the short-N characteristic method for N=1 (above) and N=2 (below). The arrows indicate the short segments through which the radiative transfer (rt) calculations will be done. The numbers above the arrows indicate the order in which rt calculations will be done on the short segments for each case. Note that the short segments will actually be in 3D space and that since we are working in cylindrical coordinates (rather than Cartesian) the relative angles in 3D space of a short segment will not be constant for different volume elements.

Fig. 2. Spatial variation of the mean intensity for a plasma on a 2DRZ grid calculated using the short-N characteristic method (open squares) for N=3. Results are shown for several heights within the cylinder. The solid lines are the analytic solutions. (above) for $\tau_R=1.825$; (below) for $\tau_R=0.165$.

Fig. 3. Spatial variation of the mean intensity for a plasma on a 2DRZ grid calculated using the short-N characteristic method (open squares) for N=20. Parameters are the same as in Figure 2.

Fig.4. Spectra produced by the long characteristic method and the short-N method for different values of the N-parameter. The spatial grid of the problem is 10x10 and a simple atomic model of 11 (eleven) energy levels for Ar is used.

Fig.5. Spectra produced by the long characteristic method and the short-N method for N=1 and N=2. The spatial grid of the problem is 10x10 and a detailed atomic model of 404 energy levels for Ar is used.

Fig.6. Spectra produced by the long characteristic method and the short-N method for different values of the N-parameter. The spatial grid of the problem is 150x80 and a simple atomic model of 11 energy levels for Ar is used.

Fig.7. Ratio of the total CPU times for post processing, with a simple atomic model of 11 energy levels of Ar, between the long characteristic method and the short-N characteristic method for N=1, 2, and 3 vs. the square root of the total number of volume elements in the spatial grid.

**Table Caption**

Table 1. Comparison of CPU times between the short-N characteristic method and the long characteristic method applied to an Ar Doped problem for a mid-spatial resolution of 50x50 and a simple atomic model for Ar of 11 energy levels. Column 1: Radiative transfer method. Column 2: N-Parameter. Column 3: Total CPU time used for post processing with *SPECT3D*. Column 4: Ratio between the total CPU time for the long characteristic method and the total CPU time for the given radiative transfer method. Column 5: CPU time used specifically for the radiative transfer calculations. Column 6: Ratio of the CPU time used specifically for radiative transfer calculations between the long characteristic method and the given radiative transfer method. Column 7: Percentage difference of the spectral intensity at the peak of the strongest line of the obtained spectrum with the given radiative transfer method, measured with respect to the value obtained through the long characteristic method.

Table 2. Comparison of CPU times between the short-N characteristic method and the long characteristic method applied to an Ar Doped problem for a mid-spatial resolution of 50x50 and a detailed atomic model for Ar of 404 energy levels. The columns are the same as in Table 1.

**Table(s)**

| Radiative Transfer Method | N-Param. | Total Time | Total Time Long / Total Time Short-N | Radiative Transfer Time | Radiative Transfer Time Long / Radiative Transfer Time Short-N | % diff (with respect to Long Charac.) |
|---|---|---|---|---|---|---|
| Long Charac. | N/A | 04h 19m 33s | 1.00 | 04h 16m 12s | 1.00 | 0.00% |
| Short-N Charac. | 1 | 10m 33s | 24.60 | 06m 17s | 48.49 | 39.34% |
| Short-N Charac. | 2 | 12m 33s | 20.68 | 07m 34s | 33.86 | 19.57% |
| Short-N Charac. | 3 | 15m 20s | 16.92 | 10m 20s | 24.79 | 10.74% |
| Short-N Charac. | 4 | 17m 33s | 14.79 | 12m 37s | 20.31 | 6.85% |
| Short-N Charac. | 5 | 23m 39s | 10.97 | 17m 51s | 14.35 | 4.82% |

Table 1.

**Table(s)**

| Radiative Transfer Method | N-Param. | Total Time | Total Time Long / Total Time Short-N | Radiative Transfer Time | Radiative Transfer Time Long / Radiative Transfer Time Short-N | % diff (with respect to Long Charac..) |
|---|---|---|---|---|---|---|
| Long Charac. | N/A | 02d 00h 02m 10s | 1.00 | 01d 15h 19m 30s | 1.00 | 0.00% |
| Short-N Charac. | 1 | 13h 24m 55s | 3.58 | 51m 52s | 45.49 | 19.84% |
| Short-N Charac. | 2 | 15h 55m 51s | 3.02 | 01h 28m 20s | 26.71 | 11.52% |
| Short-N Charac. | 3 | 16h 21m 19s | 2.94 | 01h 55m 55s | 20.35 | 6.84% |
| Short-N Charac. | 4 | 16h 24m 28s | 2.93 | 02h 20m 40s | 16.77 | 4.26% |
| Short-N Charac. | 5 | 16h 48m 00s | 2.86 | 02h 45m 23s | 14.27 | 2.90% |

Table 2.

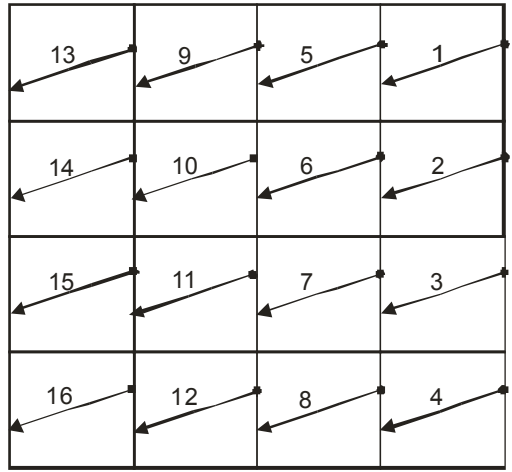

N=1

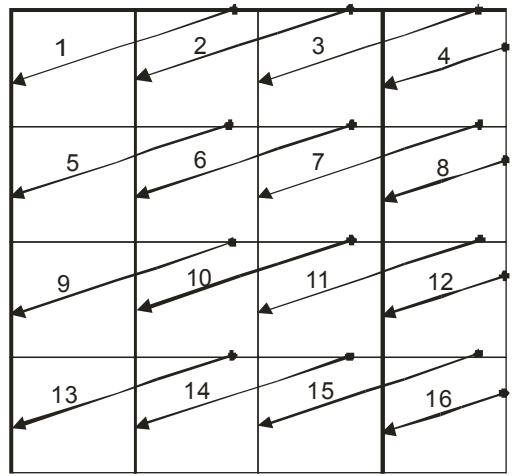

N=2

Fig. 1.

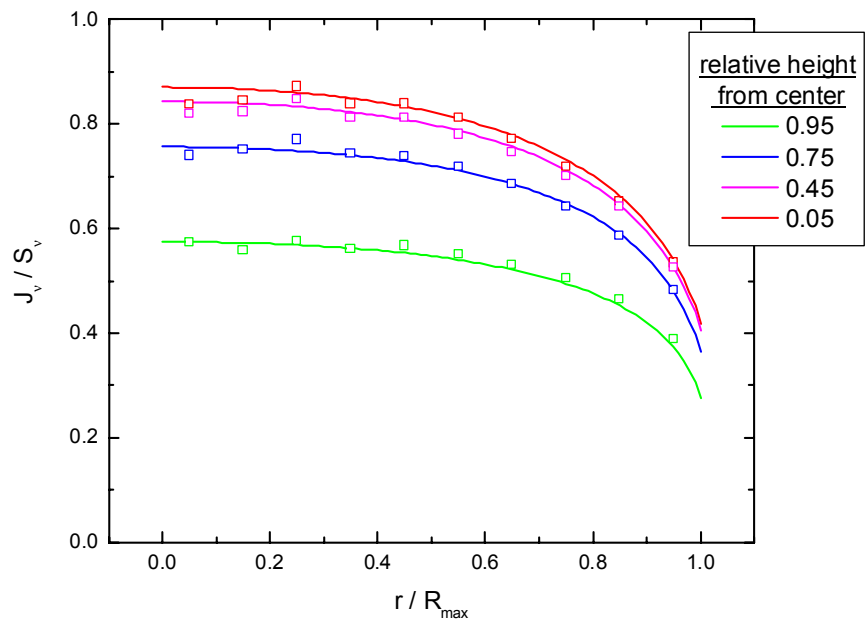

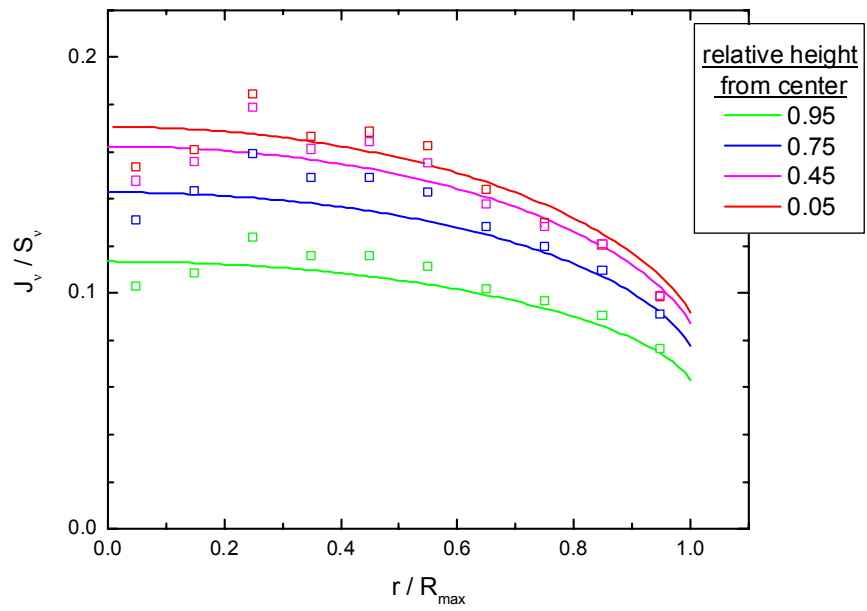

Fig. 2.

**Figure(s)**

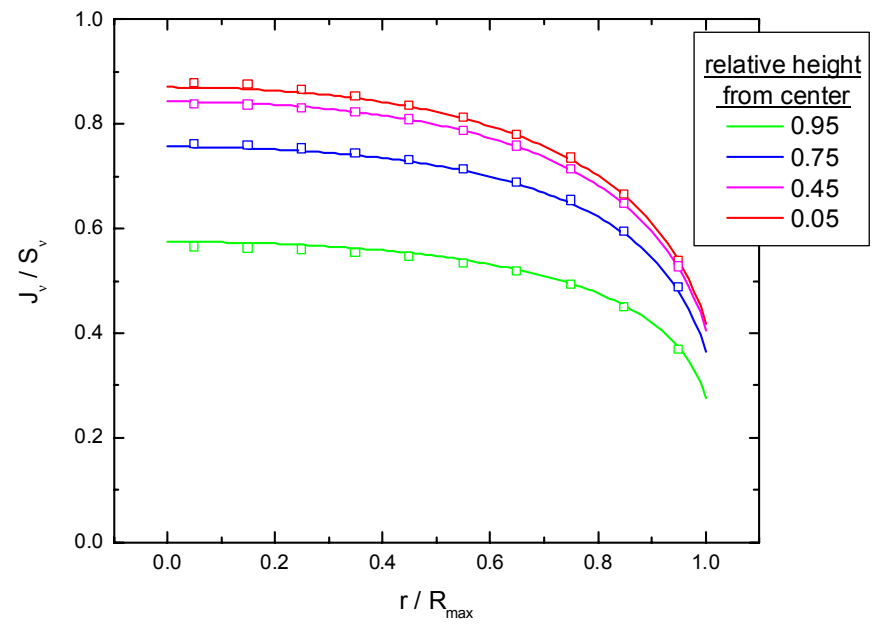

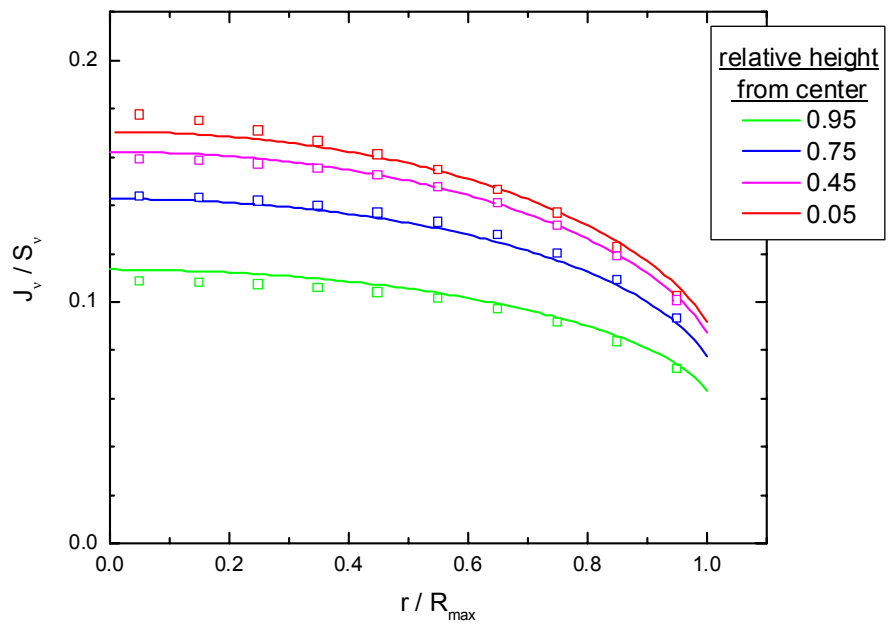

Fig. 3.

**Figure(s)**

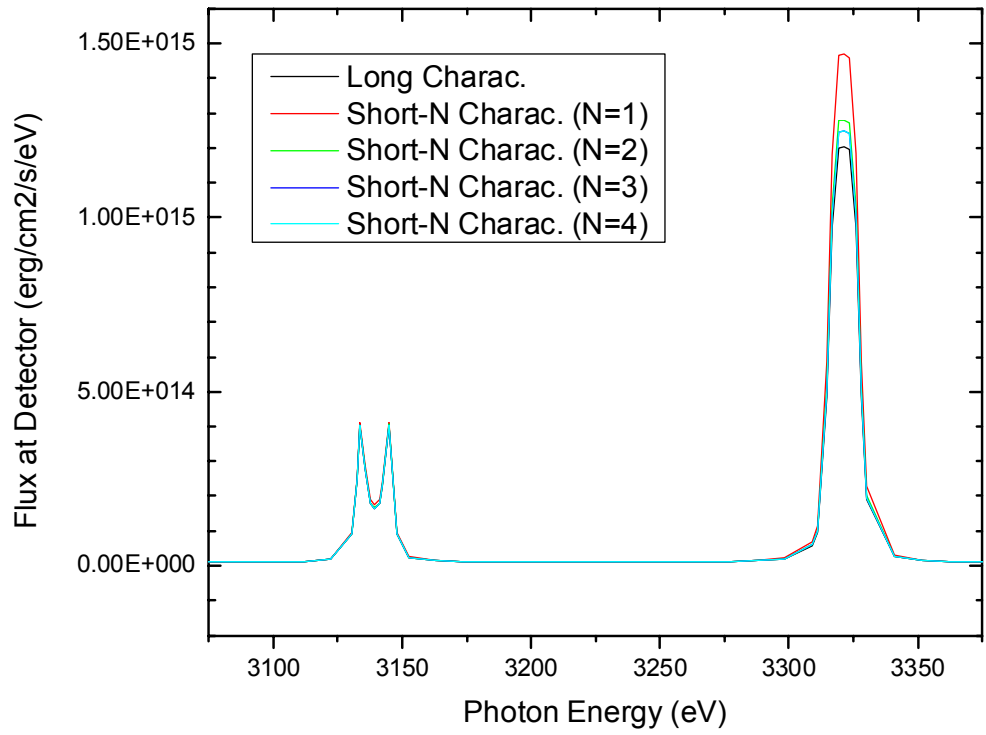

Fig. 4.

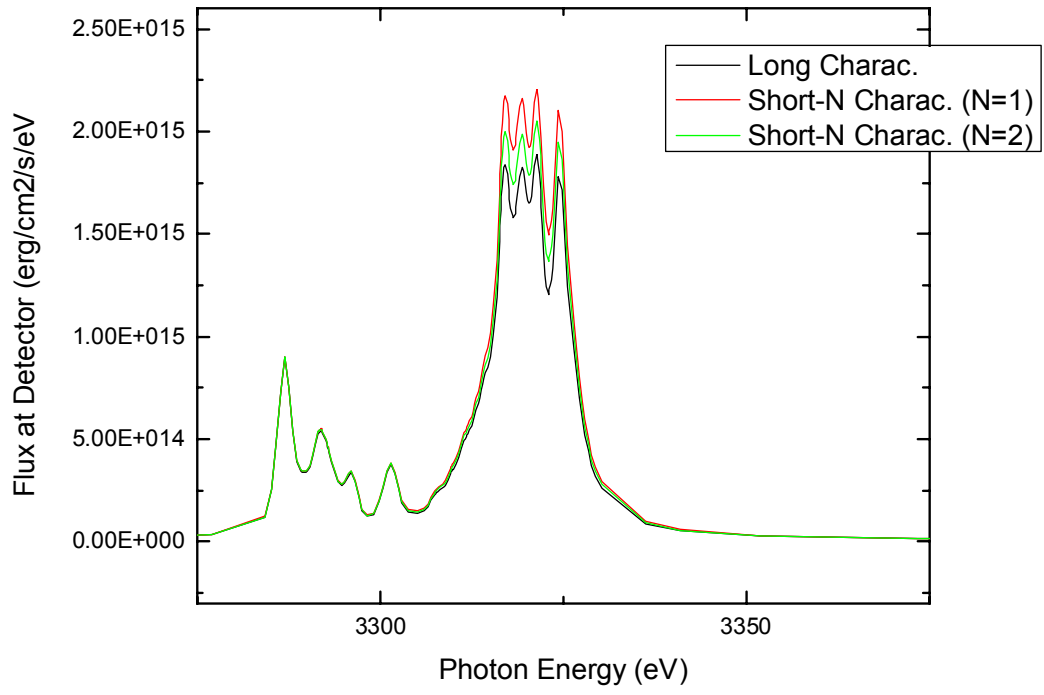

Fig. 5.

**Figure(s)**

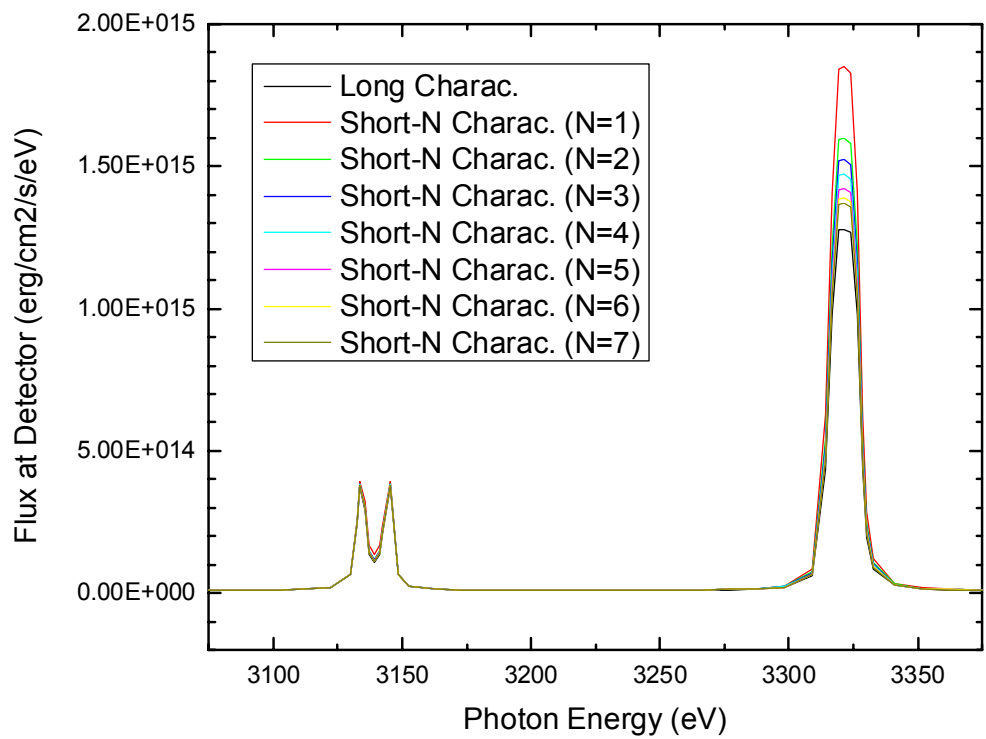

Fig. 6.

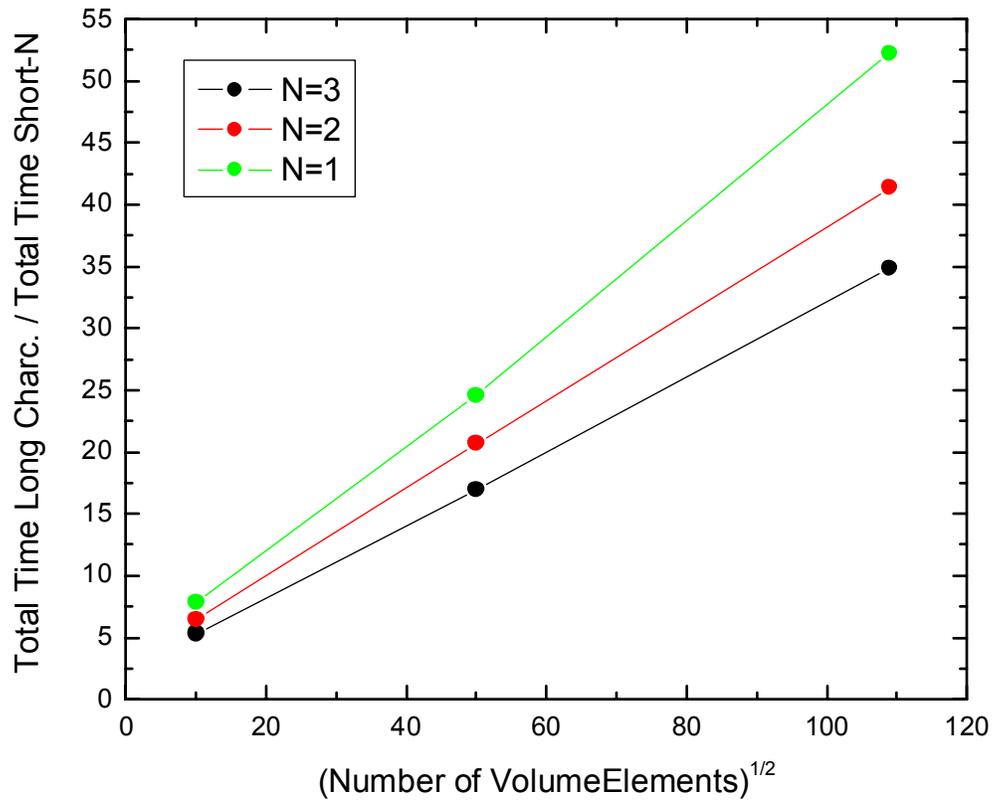

Fig. 7.